\documentclass[12pt]{article}
\usepackage{amsmath,amssymb,amsfonts,epsfig,cite,setspace,bigstrut,longtable,array,url}
\usepackage[paper=letterpaper,margin=1in]{geometry}
\usepackage{graphicx}
\usepackage{hyperref}
\usepackage[titles]{tocloft}

\newcommand{\nn}{\nonumber}

\newcommand{\comment}[1]{}

\newcommand{\cO}{{\cal O}}

\newcommand{\cF}{{\cal F}}
\newcommand{\cU}{{\cal U}}
\newcommand{\cT}{{\cal T}}

\newcommand{\IR}{\mathbb{R}}
\newcommand{\IC}{\mathbb{C}}

\newcommand{\IZ}{\mathbb{Z}}
\newcommand{\re}{{\rm~Re}}
\newcommand{\im}{{\rm~Im}}

\newcommand{\as}{\alpha(s)}
\newcommand{\at}{\alpha(t)}
\newcommand{\au}{\alpha(u)}
\newcommand{\qed}{\hfill{$\blacksquare$}}
\renewcommand{\tilde}{\widetilde}
\renewcommand{\hat}{\widehat}

\newtheorem{proposition}{\bf PROPOSITION}

\newtheorem{corollary}{\bf COROLLARY}

\newcommand{\setall}{\setcounter{equation}{0}
        \setcounter{theorem}{0}}

\renewcommand{\thefootnote}{\fnsymbol{footnote}}

\numberwithin{equation}{section}

\newcommand\be{\begin{equation}}
\newcommand\ee{\end{equation}}
\newcommand\bea{\begin{eqnarray}}
\newcommand\eea{\end{eqnarray}}
\newcommand\eref[1]{\eqref{#1}}

\begin{document}
\thispagestyle{empty}
\renewcommand{\thefootnote}{\fnsymbol{footnote}}

${}$

\vspace{2 cm}

\begin{center}

\textbf{\LARGE From Veneziano to Riemann:}\\\vspace{0.2 cm}
\textbf{\large A String Theory Statement of the Riemann Hypothesis}

\vspace{1cm}

{\large
Yang-Hui He$^{a,}$\,\footnote[1]{\texttt{hey@maths.ox.ac.uk}},
Vishnu Jejjala$^{b,}$\,\footnote[2]{\texttt{vishnu@neo.phys.wits.ac.za}},
Djordje Minic$^{c,}$\,\footnote[3]{\texttt{dminic@vt.edu}}}

\vspace{1cm}
$^{a}$\textit{Department of Mathematics, City University, London, EC1V 0HB, UK and\\
School of Physics, NanKai University, Tianjin, 300071, P.R.~China and \\
Merton College, University of Oxford, OX1 4JD, UK}\\

\vspace{0.5cm}
$^{b}$\textit{NITheP, School of Physics, and Mandelstam Institute for Theoretical Physics,\\
University of the Witwatersrand, Johannesburg, WITS 2050, South Africa}\\

\vspace{0.5cm}
$^{c}$\textit{Department of Physics, Virginia Tech, Blacksburg, VA 24061, USA}

\end{center}

\vspace{1.5cm}

\begin{abstract}
We discuss a precise relation between the Veneziano amplitude of string theory, rewritten in terms of ratios of the Riemann zeta function, and two elementary criteria for the Riemann hypothesis formulated in terms of integrals of the logarithm and the argument of the zeta function.
We also discuss how the integral criterion based on the argument of the Riemann zeta function relates to the Li criterion for the Riemann hypothesis.
We provide a new generalization of this integral criterion.
Finally, we comment on the physical interpretation of our recasting of the Riemann hypothesis in terms of the Veneziano amplitude.
\end{abstract}

\newpage
\renewcommand{\thefootnote}{\arabic{footnote}}
\setcounter{footnote}{0}

\tableofcontents

~\\
~\\
~\\

\section{String Theory and the Riemann Zeta Function}\setall
\label{secone}
Hilbert~\cite{hilbert} and P\'olya~\cite{polya} independently suggested a physical realization of the Riemann hypothesis~\cite{rh}:
if the zeroes of the zeta function in the critical strip are the eigenvalues of $\frac12 \mathbf{1} + i \mathbf{T}$ and $\mathbf{T}$ is a Hermitian operator acting on some Hilbert space, then because the eigenvalues of $\mathbf{T}$ are real, the Riemann hypothesis follows.
We do not know what the operator $\mathbf{T}$ is, nor what Hilbert space ${\cal H}$ it acts on.
Nevertheless, the Hilbert--P\'olya conjecture constitutes a tremendous insight into how a mathematics problem can be mapped to a physical system.
Many recent works seek to establish such a connection explicitly.
(See, for example,~\cite{julia, bk, khuri, connes, us, cc, sierra, LeClair} and references therein.)

In this note, we establish a physical realization of the Riemann hypothesis in terms of the properties of bosonic strings.
In particular, we consider equivalent statements of the Riemann hypothesis written as integrals of the logarithm of the zeta function or the argument of the zeta function evaluated on the critical line.
We link these expressions to the Veneziano amplitude describing the scattering of four bosonic open strings with tachyonic masses.
This opens a fascinating new connection between string theory and the physics of the Riemann zeros.
We also discuss the relation to the Li criterion for the Riemann hypothesis discussed in our previous publication~\cite{us},
and we generalize the integral criterion based on the argument of the Riemann zeta function evaluated on the critical line.

In this opening section, we point out the fundamental reason why string theory, through the form of the Veneziano amplitude, should know about the Riemann zeta function~\cite{edwards}, and thence about the Riemann hypothesis.
Our argument is based on the seminal work of Freund and Witten on the adelic strings~\cite{Freund:1987ck}.
Their work appears in the context of p-adic string theory, itself a subject of long history~\cite{Volovich:1987nq,Arefeva:1988qi,Brekke:1988dg,Brekke:1993gf}, where interesting dynamics emerges from the string worldsheet on non-Archimedean number fields.
What we study in this section is the {\it ordinary} Veneziano amplitude, however, written in a somewhat unusual form.
This expression has already appeared in~\cite{Freund:1987ck} and serves as the starting point for our investigations.
Many papers in the literature have attempted to harness the powers of the Veneziano amplitude, especially for the $p$-adic string~\cite{Arefeva:1988qi}, to address the Riemann hypothesis~\cite{links,lapidus}.
However, the relation between the Veneziano amplitude and certain elementary criteria for the Riemann hypothesis, which we point out in this paper, is new.

It is now a famous fact that the scattering amplitude~\cite{veneziano} of the bosonic string ---
when it was still known as the dual resonance model ---
is given by the Euler Beta function,
\begin{equation}
B(s,t) := \int_0^1 dx\ x^{s -1} (1-x)^{t -1} =
\frac{\Gamma(s) \Gamma(t) }{\Gamma(s + t)} ~,
\qquad \re(x) > 0 ~, \quad \re(y) > 0 ~.
\end{equation}
In particular, the four-point Veneziano amplitude is
\begin{equation}
A(s,t) = B(-\alpha(s), -\alpha(t)) ~,
\end{equation}
where $\alpha(s)$ is the Regge trajectory, usually taken as a linear function in terms of the squared four momenta (the Mandelstam variables) $s = -(k_1+k_2)^2$ and $t = -(k_2 + k_3)^2$ for external momenta $k_{i=1,2,3,4}$.
The four-point amplitude with $s$, $t$, and $u = -(k_1+k_3)^2$ symmetrized is the crossing symmetrized amplitude,
\begin{equation}\label{Astu}
A_4(s,t,u) = A(s,t) + A(t,u) + A(s,u) ~.
\end{equation}
Traditionally, one takes the linear Regge trajectory to be $\alpha(s) = 1 + \frac{1}{2}s$ and works in units where the mass squared of the tachyon is $-2$.
This implies that $s+t+u = -8$, and
\begin{equation}\label{alpha}
\alpha(s) + \alpha(t) + \alpha(u) = -1 ~.
\end{equation}
However, as we will later see, we will take a slightly different limit of the usual Regge trajectory.

In constructing a p-adic string amplitude, a useful corollary of~\cite{Freund:1987ck} is the following:
\begin{proposition}\label{FW} {\rm [Freund--Witten]}
The crossing symmetrized amplitude $A_4(s,t,u)$ in~\eqref{Astu} subject to~\eqref{alpha} can be written {\it entirely} in terms of the Riemann zeta function.
More precisely,
\[
A_4(s,t,u) = B(-\as,-\at) + B(-\at,-\au) + B(-\as,-\au) = 
\prod_{x=s,t,u} \frac{\zeta(1+\alpha(x)) }{\zeta(-\alpha(x)) } ~.
\]
\end{proposition}
\paragraph{Proof: }
We will present an expanded version of the argument in~\cite{Freund:1987ck}.
First, combining the three channels and substituting in~\eqref{alpha}, we have that
\begin{align}
\nn
A_4(s,t,u) &= 
\frac{\Gamma(-\as) \Gamma(-\at)}{\Gamma(-\as-\at)} + A(t,u) + A(s,u)
\\
\nn
&= -\frac{1}{\pi} \sin(\pi \au) \Gamma(-\au) \Gamma(-\as) \Gamma(-\at) 
+ \{ s \to t, t \to u\} + \{ s \to s, t \to u\} 
\\
\nn
&=-\frac{1}{\pi} \Gamma(-\as) \Gamma(-\at) \Gamma(-\au) \left(
\sin(\pi \as)+\sin(\pi \at)+\sin(\pi \au)
\right)
\\
\label{prod1}
&=\frac{1}{2\pi} \prod_{x=s,t,u} 2 \Gamma(-\alpha(x)) \cos\left(
\frac12 \pi \alpha(x)
\right) \ .
\end{align}

In the second line of the above, we have employed the well known {\it Euler reflection formula}:
\begin{equation}\label{reflection}
\Gamma(z) \Gamma(1-z) = \frac{\pi}{\sin{\pi z}} ~,
\end{equation}
which follows, for example, from the  Weierstra\ss\ product expansion for Gamma and sine, so that 
\begin{equation}
\Gamma(-\as-\at)^{-1} = \frac{1}{\pi} \sin\left(\pi(-\as-\at)\right) 
\Gamma(1+\as+\at) = -\frac{1}{\pi} \sin(\pi \au) \Gamma(-\au) \ .
\end{equation}
In the final line of the above, we make use of the simple but remarkable half circle trigonometric identity
\begin{equation}
\sum_{x=s,t,u} \sin(\pi \alpha(x)) = -4 \prod_{x=s,t,u} \cos\left(
\frac12 \pi \alpha(x) \right) \ ,\quad
\alpha(s) + \alpha(t) + \alpha(u) = -1 ~.
\end{equation}

Now, recall the {\it duplication formula} for the Euler Gamma function:
\begin{equation}
\Gamma(z) \Gamma(z + \frac12) = 2^{1-2z} \sqrt{\pi} \Gamma(2z) \ ,
\end{equation}
which can be rewritten as
\begin{equation}\label{dup}
\Gamma(z) = \pi^{-\frac{1}{2}} 2^{z-1} \Gamma(\frac{1}{2} z) \Gamma(\frac{1}{2} (z+1)) .
\end{equation}
Furthermore,~\eqref{reflection} can be recast, using $z \to \frac12 (z+1)$, into
\begin{equation}\label{cos}
\cos\left(
\frac12 \pi z 
\right) = 
\pi \left( \Gamma(\frac12 z + \frac12) \Gamma(-\frac12 z + \frac12) \right)^{-1}
\ .
\end{equation}
The factor $2 \Gamma(-\alpha(x)) \cos\left(
\frac12 \pi \alpha(x)
\right)$ in the product in~\eqref{prod1}, using~\eqref{dup} and~\eqref{cos}, hence becomes
\begin{align}
\nn
&2 \pi^{-\frac{1}{2}} 2^{-\alpha(x)-1} \Gamma(-\frac{1}{2} \alpha(x)) \Gamma(-\frac{1}{2} \alpha(x) + \frac12))
\pi \left( \Gamma(-\frac12 \alpha(x) + \frac12) \Gamma(\frac12 \alpha(x) + \frac12) \right)^{-1}
\\
=& 
\sqrt{\pi} 2^{-\alpha(x)} \frac{\Gamma(-\frac{1}{2} \alpha(x))}
{\Gamma(\frac12 \alpha(x) + \frac12)} \ ,
\end{align}
so that, on using~\eqref{alpha} again,
\begin{equation}\label{vene1}
A_4(s,t,u) = \frac{1}{2\pi} \prod_{x=s,t,u} \sqrt{\pi} 2^{-\alpha(x)} \frac{\Gamma(-\frac{1}{2} \alpha(x))}{\Gamma(\frac12 \alpha(x) + \frac12)} =
\sqrt{\pi} \prod_{x=s,t,u} 
\frac{\Gamma(-\frac{1}{2} \alpha(x))}{\Gamma(\frac12 \alpha(x) + \frac12)} \ .
\end{equation}
This, as the authors of~\cite{Freund:1987ck} confess, is an interesting way of writing the Veneziano amplitude.

Now, we use the famous functional equation for the Riemann zeta function
(which can be viewed as an expression of {\it duality}):
\begin{equation}
\zeta(z) = 2^z \pi^{z-1} \sin(\frac12 \pi z) \Gamma(1-z) \zeta(1-z) \ .
\label{funcid}
\end{equation}
This is valid for all complex $z$.
Multiplying both sides by $\Gamma(\frac12 z)$, again using~\eqref{reflection} to exchange $\sin(\frac12 \pi z)$ with $\pi \left( \Gamma(\frac{z}{2}) \Gamma(1 - \frac{z}{2}) \right)^{-1}$, and finally trading the $\Gamma(1-z)$ for the expression $\pi^{-\frac12} 2^{-z} \Gamma(\frac12 - \frac{z}{2}) \Gamma(1 - \frac{z}{2})$ by the duplication formula~\eqref{dup}, we have
\begin{equation}
\Gamma(\frac{1}{2} z) \zeta(z) = \pi^{z-\frac{1}{2}} \Gamma(\frac{1}{2} (1-z)) \zeta(1-z) \ ,
\end{equation}
which can be rewritten as an equality of ratios for $-z$
\begin{equation}\label{ratio}
\frac{\Gamma(-\frac12 z)}{\Gamma(\frac12z + \frac12)} = \frac{\zeta(1+z)}{\zeta(-z)} \pi^{-\frac12 - z}  \ .
\end{equation}
Upon substituting into~\eqref{vene1} and using~\eqref{alpha} we arrive at our claim.
Note that all powers of $\pi$ cancel exactly.
\qed

Several remarks are immediate.
First, the crossing symmetry is {\it necessary}.
The original unsymmetrized four-point function, written in terms of the Beta function, {\it cannot} be written purely in terms of the zeta function.
One needs to add in the full $(s,t,u)$ symmetrization of the four-point function.
The moral here is: while three Gamma functions appropriately combine into the Beta function (which was Veneziano's initial observation), the appropriate triple sum over the Beta functions gives us a ratio of pairs of Gamma functions.
It is this ratio which allows us, using the duality version~\eqref{ratio} of the functional equation, to eliminate the Gamma functions entirely in terms of ratios of zeta functions.

One might wonder, how is it possible that the Gamma functions could capture the zeros of the zeta function in~\eqref{ratio}?
In retrospect this is not surprising.
The Hadamard product (\textit{q.v.}, Section~\ref{s:V} for further discussions) for the zeta function, valid for all complex $z$, is
\begin{equation}
\zeta(z) = \pi^{\frac{z}{2}} \frac{\prod_\rho \left(1 - \frac{z}{\rho}\right)}{2(z-1)\Gamma(1 + \frac{z}{2})} \ , \label{had}
\end{equation}
for non-trivial zeros $\rho$.
The denominator encodes the simple pole of the zeta function at $z=1$ and the trivial zeros at the negative even integers.
If $z$ is a non-trivial zero, the numerator vanishes.
Using the fact that the Gamma function is the generalized factorial function so that $\Gamma(1+x) = x\Gamma(x)$, we may use~\eqref{had} to write the ratios of the zeta functions as
\begin{equation}
\frac{\zeta(1+z)}{\zeta(-z)} =
\frac{{\pi^{\frac12 + z}} \Gamma(-\frac12 z)}{\Gamma(\frac12z + \frac12)}
\prod_\rho \frac{\rho - 1 - z}{\rho + z} \ .
\label{hadprod}
\end{equation}
However, as a consequence of the functional equation of the zeta function and independent of the Riemann hypothesis, $\rho$ is a root if and only if $1 - \rho$ is.
Thus, the product over the non-trivial zeros cancels exactly because they pair up telescopically.
In a way, this is an alternative derivation of~\eqref{ratio} using the Hadamard product.
Because the product is unity,~\eqref{hadprod} has the same content as~\eqref{ratio}.
In sum, the ratio of zeta functions in~\eqref{ratio} does {\it not} know about the non-trivial zeros, only the trivial ones coming from the Gamma function.

Given this fact, what could the Freund--Witten rewriting of the Veneziano amplitude teach us about the Riemann zeros?
In their paper, Freund and Witten concentrate on the factorization of the four-point Veneziano amplitude in terms of an infinite product of {\it primes $p$} of the corresponding p-adic Veneziano amplitudes.
(This factorization {\it does not} occur for higher order amplitudes~\cite{Brekke:1993gf}.)
Nevertheless, even though the non-trivial zeros cancel out to leave the usual spectrum of string theory, the string might know about the Riemann zeros and thus about the distribution of primes in some roundabout way. 
For example, the string theory excitations, governed by the usual zeros of the Euler gamma function, or the trivial Riemann zeros, might be expressed in a dual way in terms of non-trivial Riemann zeros.
As string theory is a unitary theory, this dual rewriting should also be described by a unitary theory, and thus, in the dual rewriting, string theory might represent the Hilbert--P\'olya physical system whose excitations are given by the non-trivial Riemann zeros.
This is largely conjectural, however.

In what follows we achieve a more modest outcome.
In Section~\ref{equivalences}, we concentrate not on the spectrum of the Veneziano amplitude rewritten in the Freund--Witten form, but on its
{\it argument}.
We clarify our intuition by formulating equivalent physical criteria for the Riemann hypothesis in terms of integrals involving the zeta function that find a natural expression in the properties of the Veneziano amplitude (more specifically, its argument).
The principal new result in this paper is the derivation of Proposition~\ref{vv}. In Section 3, we provide a generalization of the
integral criterion based on the argument of the Riemann zeta function evaluated on the critical line and we also point out the relation to the Li criterion for the Riemann hypothesis.
In Section~\ref{physical}, we offer a physics perspective on the Volchkov--Veneziano criterion that we obtain.
\section{Statements Equivalent to the Riemann Hypothesis}\label{equivalences}\setall
It was shown in~\cite{volchkov} that the Riemann hypothesis is {\it equivalent} to the exact integral expression
\begin{equation}\label{volchkov1}
\int_0^\infty dT\ \frac{1-12T^2}{(1+4T^2)^3} \int_{\frac12}^\infty d\sigma\ \log \left| \zeta(\sigma + i T) \right|
= \pi \frac{3 - \gamma}{32} ~,
\end{equation}
where $\gamma$ is the Euler--Mascheroni constant.
Since this remarkable criterion, which was discovered only in the 1990s, several equivalent formulations of the Riemann hypothesis in terms of integrals of logarithms of the zeta function have been found.
A good summary is~\cite{log}, which includes as special cases, inter alia, the following two criteria, one involving the log absolute value of the zeta function on the critical line and the other the argument of the zeta function on the critical line.

\begin{itemize}
\item
The Balazard--Saias--Yor~\cite{BSY} criterion: The Riemann hypothesis is equivalent to 
\begin{equation}\label{BSY}
\int_{-\infty}^\infty dT\ \frac{\log \left| \zeta(\frac12 + i T) \right|}{\frac14 + T^2} = 0 \ .
\end{equation}

\item
The Volchkov~\cite{volchkov} criterion: The Riemann hypothesis is equivalent to
\begin{equation}\label{volchkov}
\int_0^\infty dT\ \frac{2 T \arg \zeta(\frac12 + i T)}{(\frac14 + T^2)^2}
=
\pi(\gamma - 3) ~.
\end{equation}
\end{itemize}
Now, as presented, this second criterion is different in form from the original statement~\cite{volchkov}, and it was generalized in~\cite{log}. 
We will not make use of the statement cast in~\eqref{volchkov1}, but, instead, the one in~\eqref{volchkov};
it is expedient to discuss these criteria in more detail, but we will leave this to Section~\ref{s:V}.

Can any of these criteria be rephrased as a statement about the Veneziano amplitude?
From Proposition~\ref{FW}, we write explicitly
\begin{equation}
A_4(s,t,u) = 
\frac{\zeta(1+\alpha(s)) }{\zeta(-\alpha(s))}
\frac{\zeta(1+\alpha(t)) }{\zeta(-\alpha(t))}
\frac{\zeta(-\as -\at) }{\zeta(1+\as+\at)}
\ ,
\end{equation}
where we recall that $\as+\at+\au=-1$.
Due to this constraint, we will sometimes drop that last argument $u$ from the symmetrized four-point amplitude $A_4$.

The Balazard--Saias--Yor result does not yield anything immediately useful but nevertheless does produce an interesting integral which we derive in Appendix~\ref{ap:BSY}.
It turns out, however, that a direct manipulation can recast the Riemann hypothesis in terms of properties of the Veneziano amplitude using the Volchkov criterion.
Two limits lead to particularly striking expressions for the open string four tachyon scattering amplitude:
\begin{eqnarray}
\mathbf{(A)} &\qquad& \as = \at = - \frac12 + i T \ , \quad T \in \IR \label{venlimit} ~, \\
\mathbf{(B)} &\qquad& \as = -iT ~, \quad \at = -\frac12 \ , \quad T \in \IR \label{newlimit} ~.
\end{eqnarray}

\subsection{Veneziano and Volchkov}\label{volven}
As we demonstrate in Appendix~\ref{ap:BSY}, considering only the log absolute value of the zeta function on the critical line loses precious information about the non-trivial zeros.
This is a consequence of the fact that $\zeta(\bar z) = \overline{\zeta(z)}$.
Consequently, as we have noted above, the Balazard--Saias--Yor criterion is not helpful for our purposes.
Likewise, considering case {\bf A} as discussed in \eqref{venlimit} loses information about the critical zeros as we explain in Appendix~\ref{ap:VV} in pedagogical detail.

We therefore turn now to case \textbf{B}, \eqref{newlimit},
from which we write
\be
A_4(\alpha^{-1}(-iT),\alpha^{-1}(-\frac12)) = \frac{\zeta(1-iT)}{\zeta(iT)} \frac{\zeta(\frac12+iT)}{\zeta(\frac12-iT)} = \sqrt\pi\, \frac{\Gamma(\frac{i}{2}T)}{\Gamma(\frac12-\frac{i}{2}T)} \frac{\Gamma(\frac14-\frac{i}{2}T)}{\Gamma(\frac14+\frac{i}{2}T)} ~.
\ee
Crucially, we know that $\zeta(iT)$ and $\zeta(1-iT)$ have no zeros\footnote{
The classic result of Hadamard and de la Vall\'ee Poussin from 1896 that there are no zeros of the zeta function on the boundary of the critical strip is central to the proof of the Prime Number Theorem.}
for any $T\in \mathbb{R}$, and hence, the ratio of the two is entire.
This permits us to define a dressed version of the Veneziano amplitude:
\begin{equation}
\mbox{\fbox{$
\hat{A}(T) := A_4(\alpha^{-1}(-iT),\alpha^{-1}(-\frac12)) \frac{\zeta(iT)}{\zeta(1-iT)}$}} ~.
\end{equation}
Since
\be
\arg\hat{A}(T) = 2\arg\zeta(\frac12+iT) ~,
\ee
and, in particular, because there is no correction factor $2\pi\hat{n}(T)$, we have:
\begin{proposition}\label{vv}
An equivalent statement of the Riemann hypothesis in terms of the Veneziano amplitude is that
\begin{equation}
\int_{0}^\infty dT\ 
\frac{2 T\ {\arg \hat{A}(T)}}{(\frac14 + T^2)^2}
= 2\pi ( \gamma - 3)
\ . 
\label{prop}
\end{equation}
\end{proposition}
In the above we recall that the Veneziano amplitude is the symmetrized four-point function 
$A_4(s,t,u) = B(-\as,-\at) + B(-\at,-\au) + B(-\as,-\au) = 
\prod\limits_{x=s,t,u} \frac{\zeta(1+\alpha(x)) }{\zeta(-\alpha(x)) }$ with the constraint that $\as + \at + \au = -1$.
We consider this function when $\as = -iT$ and $\at = -\frac12$.
More succinctly, we can rewrite the above criterion as
\begin{equation}\label{vv2}
\exp\left\{ i \int_{0}^\infty d\left( \frac{1}{\frac14 + T^2}\right) \arg \hat{A}(T) \right\}
= e^{-2\pi i \gamma} \simeq -0.884600 + 0.466350\, i \ .
\end{equation}
We observe for later that $\arg \hat{A}(T)$ is an odd function of $T$.

We call Proposition~\ref{vv} the \textit{Volchkov--Veneziano criterion} for the Riemann hypothesis.
The integral~\eqref{prop} holds if and only if the Riemann hypothesis is true.

We can also state the relation between the Riemann hypothesis and other string amplitudes.
For example, the scattering of four closed string tachyons is described by the Virasoro--Shapiro amplitude~\cite{vs}.
As we discuss in Appendix~\ref{ap:VS}, we may substitute this in Proposition~\ref{vv}.

\section{The Volchkov Criterion Revisited}\label{s:V}
As promised, we now examine the Volchkov criterion in further detail, paying attention to the nuances in its proof and in due course find unexpected connections with our earlier work~\cite{us}.
Let us begin with a more detailed discussion on the Hadamard product representation for the zeta function, along the lines discussed in Chapter~12 of~\cite{daven}, which constitutes a basic ingredient of~\cite{volchkov,log}.

Recall the Riemann xi function
\begin{equation}
\xi(z) = \frac12 z (z-1) \pi^{-\frac{z}{2}} \Gamma(\frac12 z) \zeta(z) \ ,
\end{equation}
which is constructed --- the poles of Gamma canceling the trivial zeros of zeta at all negative even integers --- to have zeros $\rho = \sigma + i T$ only in the critical strip $\mathfrak{S} = \{z \ | \ 0 \le \re(z) \le 1\}$.
It therefore has Weierstra\ss\ expansion
\begin{equation}
\xi(z) = \exp(A + B z) \prod_{\rho \in \mathfrak{S}} 
\left(1 - \frac{z}{\rho} \right) \exp(\frac{z}{\rho})
\end{equation}
for constants $A$ and $B$ to be determined.

Taking the logarithmic derivative readily gives
\begin{align}
\nn
\frac{\xi'(z)}{\xi(z)} & = B + \sum_{\rho \in \mathfrak{S}} 
\left(\frac{1}{z - \rho} + \frac{1}{\rho}  \right) \\
\label{logD}
\frac{\zeta'(z)}{\zeta(z)} & = B - \frac{1}{s-1} + \frac12 \log \pi -
  \frac12 \frac{\Gamma'(\frac12 z + 1)}{\Gamma(\frac12 z +1)} + 
\sum_{\rho \in \mathfrak{S}} 
\left(\frac{1}{z - \rho} + \frac{1}{\rho}  \right) \ .
\end{align}
Explicitly, we can see the trivial zeros in the zeta function at $z = -2,-4,\ldots$ because the logarithmic derivative of the Gamma function, using its Weierstra\ss\ product is
\begin{equation}\label{logDGamma}
-\frac12 \frac{\Gamma'(\frac12 z + 1)}{\Gamma(\frac12 z +1)} =
\frac12 \gamma + \sum_{n=1}^\infty \left(
\frac{1}{z +2n} - \frac{1}{2n} 
\right) \ .
\end{equation}

The constants $A$ and $B$ can be fixed by evaluation:
\begin{equation}\label{AB}
A = \log \zeta(0) = -\log2 \ , \quad
B = - \frac{\xi'(1)}{\xi(1)} = \frac{\xi'(0)}{\xi(0)} =
-\frac12 \gamma - 1 +\frac12 \log 4 \pi \ .
\end{equation}

Crucially, the constant $B$ is also a vital sum.
The functional equation
\begin{equation}\label{func}
\xi(z) = \xi(1-z)
\end{equation}
applied to the logarithmic derivative~\eqref{logD} implies that
\begin{equation}
B + \sum_{\rho \in \mathfrak{S}} 
\left(\frac{1}{1-z - \rho} + \frac{1}{\rho}  \right)
=
- B - \sum_{\rho \in \mathfrak{S}} 
\left(\frac{1}{z - \rho} + \frac{1}{\rho}  \right) \ .
\end{equation}
Remembering that again by the functional equation $\rho$ is a zero if and only if $1 - \rho$ is, the fractions involving $z$ cancel.
Then
\begin{equation}\label{B}
B = - \sum_{\rho \in \mathfrak{S}} \frac{1}{\rho} = 
- \sum_{\rho \in \mathfrak{S}, T>0} 
   \frac{1}{\rho} + \frac{1}{\overline{\rho}}= 
-2 \sum_{\rho \in \mathfrak{S},T > 0} \frac{\sigma}{\sigma^2 + T^2} 
\simeq -0.023
\ .
\end{equation}
These together give us the nice compact expansion of the Hadamard product:
\begin{equation}\label{hadamard}
\xi(z) = \frac12 \prod_{\rho \in \mathfrak{S}} \left(
1 - \frac{z}{\rho}
\right) \ . 
\end{equation}

Prepared with these rudiments, let us follow the arguments of~\cite{log} in the derivation of~\eqref{volchkov}.
From contour integration of~\eqref{logD}, we discover that $N(x) = \frac{1}{\pi} \im\log \xi(\frac12+ix)$.
Earlier, we encountered the exact expression in~\eqref{countingfn}.
For any integrable smooth function $G(x)$ which decays faster than $N(x)$ at $T \to \infty$, we have that
\begin{equation}\label{S}
S := \sum_{\rho \in \mathfrak{S},T > 0} G(T) = \int_0^\infty  d N(x)\ G(x)
= -\int_0^\infty dx\ G'(x) N(x)
\ .
\end{equation}
Let us briefly remember that as $N(x)$ counts the numbers of zeros in the critical strip up to a height $x$, $dN(x)$ in the previous equation is a sum of delta functions corresponding to each of the zeros encountered as we move away from the real axis.
Now, simply take
\begin{equation}
G(x) = (\frac14 + x^2)^{-1} \ , 
\qquad -G'(x)=\frac{2x}{(\frac14+x^2)^2} ~,
\end{equation}
and we have that
\begin{equation}\label{SB}
S = \sum_{\rho \in \mathfrak{S},T > 0} \frac{1}{\frac14 + T^2} 
\qquad \mbox{ iff RH } \qquad  = 
2\sum_{\rho \in \mathfrak{S},T > 0} \frac{\sigma}{\sigma^2 + T^2} 
= -B \ .
\end{equation}
In the above, one uses the fact that the function $f_T(\sigma) = \frac{\sigma}{\sigma^2+T^2}$ is convex (for any $T> 1$) in the interval $[0,1]$ so that $f_T(\sigma) + f_T(1-\sigma) \le 2 f_T(\frac12)$ for all $\sigma \in [0,1]$ and with equality holding iff $\sigma = \frac12$.
On the other hand, we recall from~\eqref{B} that $B = -2 \sum_{\rho \in \mathfrak{S},T > 0} f_T(\sigma)$ {\it unconditionally} upon the Riemann hypothesis.
Hence, if and only if the Riemann hypothesis holds and $\sigma = \frac12$ can we equate the two sums.
We emphasize this relation: that the sum $S$, which as we saw can be found via integrating our chosen function $G(x)$ against $dN(x)$, should equal to our constant $-B$ is equivalent to the Riemann hypothesis.

We can now integrate term by term to evaluate the sum $S$. First, we have that
\begin{equation}
\int_0^\infty dx\ \frac{-2x}{(\frac14 + x^2)^{2}} (1 - \frac{\log\pi}{2\pi} x) = \frac12 \log\pi - 4 \ .
\end{equation}
Next, we have the second term, by parts:
\begin{equation}
\frac{1}{\pi} \int_0^\infty dx\ G'(x) \im \log\Gamma(\frac14+\frac{i}{2}x) = 
\frac{1}{\pi} \re \int_0^\infty dx\ G(x) 
\frac{-\frac{1}{2}\Gamma'(\frac14+\frac{i}{2}x)}{\Gamma(\frac14+\frac{i}{2}x)}
\ .
\end{equation}
This last expression can be evaluated using~\eqref{logDGamma}, and one finds that this is $\frac\gamma2 + \log 2$.
Thus, combining all terms and using the explicit expression for $B$, we arrive at
\begin{equation}
S = -B = -(\frac12\log \pi - 4) -(\frac\gamma2+ \log2) - \frac{1}{\pi} \int_0^\infty dx\ G'(x) \arg \zeta(\frac12 + ix) \ ,
\end{equation} 
being equivalent to the Riemann hypothesis, or that
\begin{equation}
\int_0^\infty dT\ \frac{2 T \arg \zeta(\frac12 + i T)}{(\frac14 + T^2)^2}
=
\pi(\gamma - 3) \ .
\end{equation}

\subsection{Relation to Li's Criterion}
We see that the crucial step in Volchkov was the integration against $N(x)$; this, together with the emergence of $\sum_{\rho \in \mathfrak{S}} \frac{1}{\rho} \simeq 0.023$, should be reminiscent of a criterion contemporary thereto, \textit{viz.}, the criterion of Li~\cite{li}, which was recently shown to admit an analogous integral representation~\cite{us}.

Let us recall Li's criterion.
\begin{proposition}\label{li} {\rm [Li]}
Consider the Taylor expansion
\[
\log \xi(\frac{1}{1-z}) = -\log 2 + \sum_{n=1}^\infty \frac{k_n}{n} z^n
\ ,
\]
then the Riemann hypothesis is equivalent to the positivity of the coefficients\footnote{
We will discuss these equivalent representations of $k_n$ in Appendix~\ref{ap:Li}.
}
\[
k_n = \sum_{\rho \in \mathfrak{S}} 
\left[1 - \left(1 - \frac{1}{\rho}\right)^n \right] 
= \frac{1}{(n-1)!} \frac{d^n}{dz^n} \left[z^{n-1} \log \xi(z) \right]_{z=1}
> 0 \ ,
\forall \ n = 1,2,3,\ldots
\ .
\]
\end{proposition}

In~\cite{us}, we obtained an integral representation of the Li coefficients:
\begin{equation}\label{chu}
k_n = n \int_0^\infty dx\ \frac{2x}{(\frac14 + x^2)^{2}} N(x) 
\,  \cU_{n-1}\left( \frac{x^2 - \frac14}{x^2 + \frac14} \right)
\ ,
\end{equation}
where $\cU_{n-1}(\cos \theta) = \sin(n \theta) / \sin(\theta)$ is the Chebyshev polynomial of the second kind.
Hence, Volchkov's integral is the case $n=1$ for which $\cU_0(\frac{x^2 - \frac14}{x^2 + \frac14})=1$.
In other words, adhering to the notation in~\eqref{S}
\begin{equation}
S = -B = k_1 \ ,
\end{equation}
and the Volchkov criterion is equivalent to the statement that
\begin{equation}
\frac{1}{\pi}\int_0^\infty dx\ \frac{2 x \arg \zeta(\frac12 + i x)}{(\frac14 + x^2)^2} = k_1 - 4 + \frac{\gamma}{2} + \log(2 \sqrt{\pi}) \ .
\end{equation}

\subsection{Generalizing the Volchkov Criterion}
Can we now imitate Volchkov's method and generate analogous statements for all the $n>1$?
We will shortly see that the validity of this integral representation, generalizing the case of $n=1$ of Volchkov, is equivalent to the Riemann hypothesis.

Indeed, the form of the integration kernel in~\eqref{chu} suggests that we should take
\begin{equation}\label{Gn}
-{G}_n'(x) = n\, \cU_{n-1}\left( \frac{x^2 - \frac14}{x^2 + \frac14} \right)
\frac{2x}{(\frac14 + x^2)^2} \ ,
\qquad
{G}_n(x) = 
2 - 2 \cT_n\left( \frac{x^2 - \frac14}{x^2 + \frac14} \right) \ ,
\end{equation}
where $\cT_n(\cos \theta) := \cos(n \theta)$ 
is the Chebyshev polynomial of the first kind.
Note that $G_{n=1} = (\frac14 + x^4)^{-1}$ is our previous case.
The integration constant $2$ is added for convergence and also plays a crucial r\^ole in the positivity criterion as we shall now see.
Incidentally, this gives the nice compact expression for our integral formula for the Li coefficients noted, but not emphasized, in~\cite{us}:
\begin{equation}
k_n = 2 \int_0^\infty d \cT_{n}\left( \frac{x^2 - \frac14}{x^2 + \frac14} \right)\ N(x) \, 
=
\sum_{\rho \in \mathfrak{S}, T>0} \left(
2 - 2 \cT_n\left( \frac{T^2 - \frac14}{T^2 + \frac14} \right) \right)
\ .
\end{equation}
Again, this is contingent on the Riemann hypothesis.

The reason that the integral representation for $k_n$ is valid on the Riemann hypothesis, or indeed, if and only if the Riemann hypothesis, is as follows.
The comparison, in light of~\eqref{SB}, is
\begin{equation}\label{box}
\begin{array}{ccc}
\mbox{\fbox{$
\begin{array}{rcl}
&&\sum_{\rho \in \mathfrak{S},T > 0} G_n(T) 
\\
&=& \int_0^\infty  d N(x)\ G_n(x)
\\
&=& n \int_0^\infty dx\ \frac{2x}{(\frac14 + x^2)^{2}} N(x) 
  \ \cU_{n-1}\left( \frac{x^2 - \frac14}{x^2 + \frac14} \right)
\\
&=& \sum_{\rho \in \mathfrak{S}, T>0} 
\left( 2 - 2 \cT_n\left( \frac{T^2 - \frac14}{T^2 + \frac14} \right) \right)
\end{array}
$}}
\stackrel{?}{\longleftrightarrow}
\mbox{\fbox{$
\begin{array}{rcl}
k_n &=&
\frac{1}{(n-1)!} \frac{d^n}{dz^n} \left[z^{n-1} \log \xi(z) \right]_{z=1}
\\
&=&\sum_{\rho \in \mathfrak{S}} 
\left[1 - \left(1 - \frac{1}{\rho}\right)^n \right]
\end{array}
$}}
\end{array}
\ .
\end{equation}
Now, the right of the bidirectional arrow with the question mark, being equaling to $k_n$ is {\it unconditional} upon the Riemann hypothesis because we are summing over all zeros in the critical strip.
The left hand side, on the other hand, is equal to the sum of the Chebyshev polynomials over the {\it imaginary} part of the critical zeros, \textit{viz.}, $\sum_{\rho \in \mathfrak{S}, T>0} 
\left( 2 - 2 \cT_n\left( \frac{T^2 - \frac14}{T^2 + \frac14} \right) \right)$ and knows nothing about the {\it real} part $\sigma$.
If the Riemann hypothesis holds and the $\stackrel{?}{\longleftrightarrow}$ can be replaced by an equal sign is obvious: simply take $\sigma = \frac12$.
That the converse direction should hold and the replacement of $\stackrel{?}{\longleftrightarrow}$ by equality would imply Riemann hypothesis is less obvious but proceeds along the same lines as Volchkov.
Consider the sum on the right
\begin{equation}
S = \sum_{\rho \in \mathfrak{S}} 
\left[1 - \left(1 - \frac{1}{\rho}\right)^n \right]
= \sum_{\rho \in \mathfrak{S},T>0} 
\left[1 - \left(1 - \frac{1}{\sigma+i T}\right)^n 
+ 1 - \left(1 - \frac{1}{\sigma - iT}\right)^n 
\right] 
\end{equation}
which is true simply because of the reflection principle and that all zeros of the xi function appear as conjugate pairs in the critical strip.
This is in turn equal to
\begin{align}
\nn
S=&\sum_{\rho \in \mathfrak{S},T>0} \left[ 2 - 
\left[
\left( \frac{\sigma-1 + iT}{\sigma + iT} \right)^n
+
\left( \frac{\sigma-1 - iT}{\sigma - iT} \right)^n
\right] \right]
\\
\nn
=&
\sum_{\rho \in \mathfrak{S},T>0}  \left[ 2 - 
\left(
1 + \frac{1-2\sigma}{\sigma^2+T^2}
\right)^{\frac{n}{2}}
\left(
\exp\left(i n \arg[1+\frac{1}{-\sigma+i T}] \right)
+
\exp\left(i n \arg[1-\frac{1}{\sigma+i T}] \right)
\right) \right]
\\
\nn
=&
\sum_{\rho \in \mathfrak{S},T>0} \left[ 2 - 
2 \left(\frac{(\sigma-1)^2+T^2}{\sigma^2+T^2}\right)^{\frac{n}{2}}
\cos\left(
n \cos^{-1} \frac{\sigma(\sigma-1) + T^2}{\sqrt{(\sigma^2+T^2)((\sigma-1)^2 + T^2)}}
\right) \right]
\\
=&
\sum_{\rho \in \mathfrak{S},T>0}  \left[ 2 - 
2 \left(\frac{(\sigma-1)^2+T^2}{\sigma^2+T^2}\right)^{\frac{n}{2}}
\cT_n\left(
\frac{\sigma(\sigma-1) + T^2}{\sqrt{(\sigma^2+T^2)((\sigma-1)^2 + T^2)}}
\right) \right]
\ ,
\end{align}
where we recalled the definition of the Chebyshev polynomial of the first kind in the last step.
If the sum is equal to its value at $\sigma = \frac12$ {\it only} at $\sigma = \frac12$, which is the sum on in the bottom of the left hand side box in~\eqref{box}, then we would have the Riemann hypothesis.
Indeed, this is in congruence with the fact that if indeed the left hand side is a correct representation of $k_n$, then since $\cT_n(x)$ is a cosine function of a real input it must be less than or equal to one so that each term in the summand on the left hand side is positive, and we have Li's positivity criterion.

The problematic term is, of course, the pre-factor $\left(\frac{(\sigma-1)^2+T^2}{\sigma^2+T^2}\right)^{\frac{n}{2}}$, which could exceed one for some hypothetically large $T$ whose corresponding $\sigma \ne \frac12$.
It suffices to show that this is unity {\it if and only if} for $\sigma = \frac12$.
Indeed, for any $T>0$ and $n \in \IZ_+$, this is clearly true.
Therefore $f_{n,T}(\sigma) = f_{n,T}(\frac12) = 2 - 2\cT_n\left( \frac{T^2 - \frac14}{T^2 + \frac14} \right)$, \textit{i.e.}, if the right box in~\eqref{box} is equal to the left box, then $\sigma$ can only equal to $\frac12$, and the Riemann hypothesis holds.

Thus, the generalization of Volchkov's method is to perform the integral
\begin{equation}
\sum_{\rho \in \mathfrak{S}, T>0}
{G}_n(T) =  \left. \int_0^\infty  d N(x)\ {G}_n(x)
= N(x) G_n(x) \right|_0^\infty -\int_0^\infty dx\ {G}_n'(x) N(x)
\ .
\end{equation}
As the first zero in the critical strip is approximately $\frac12 + 14.13472514\, i$, $N(0)$ equals zero.
The asymptotic expression for the number of zeros with imaginary parts between $0$ and $T$ is
\begin{equation}
N(T) \approx \frac{1}{\pi} \arg \xi(\frac12 + iT) = \frac{T}{2\pi}\log \frac{T}{2\pi} - \frac{T}{2\pi} + \frac78 + {\cal O}(\frac1T) ~.
\end{equation}
For large $x$, $G_n(x) \simeq n^2 x^{-2} + {\cal O}(x^{-4})$, which means
\begin{equation}
\lim_{x\to\infty} G_n(x) N(x) = \frac{n^2\log x}{2\pi x} - \frac{n^2(1+\log2\pi)}{2\pi x} + {\cal O}(\frac{1}{x^2}) \longrightarrow 0 ~.
\label{bdypiece}
\end{equation}
Thus, when we integrate by parts, we may drop the boundary term.

Again, we break the integral into the ``average'' (\textit{viz.}, $I_1 + I_2$) and ``oscillatory'' (\textit{viz.}, $I_3$) parts:
\begin{align}
\nn
I_1(n) & := -\int_0^\infty dx\ {G}'_n(x) \left(1 - \frac{\log\pi}{2\pi} x\right) 
\ ;
\\
\nn
I_2(n) & := -\int_0^\infty dx\ {G}'_n(x) \left(\frac{1}{\pi} \im \log \Gamma(\frac14 + \frac{i}{2} x)\right) \ ;
\\
I_3(n) & := -\int_0^\infty dx\ {G}'_n(x) \left( \frac{1}{\pi} \arg \zeta(\frac12 + i x) \right) \ .
\end{align}
As always, the prime denotes differentiation by $x$.

To evaluate these integrals, it is expedient to use the generating function for the Chebyshev polynomials, \textit{viz.},
\begin{equation}\label{genG}
\sum_{n=0}^\infty \cT_n(x) t^n = \frac{1-tx}{1-2tx+t^2}
\quad
\Longrightarrow
\quad
G(x,t) = \sum_{n=0}^\infty G_n(x) t^n = 
\frac{4 t (1+t)}{4 (1-t)^3 x^2+(1+t)^2 (1-t)}
\ .
\end{equation}

The evaluation of $I_1(n)$ is elementary.
The generating function~\eqref{genG} readily gives that of $G'(x)$ upon differentiation.
Hence,
\begin{align}
\nn
\sum_{n=0}^\infty I_1(n) t^n &=
\int_0^\infty dx\ 
\left(-\frac{d}{dx} \frac{4 t (1+t)}{4 (1-t)^3 x^2+(1+t)^2 (1-t)} \right) \left(1 - \frac{\log\pi}{2\pi} x\right)
\\ 
\nn
&= -\frac{t^2(8+\log\pi)-t(8-\log\pi)}{2 (1-t)^2 (1+t)}
= \sum_{n=1}^\infty \left(
-\frac{n}{2} \log \pi +2-2 (-1)^n
\right) t^n
\\
\Longrightarrow
& \quad
I_1(n) = - \frac{n}{2} \log \pi + 2(1 - (-1)^n)~.
\label{I1}
\end{align}

Next, let us derive an expression for $I_2(n)$.
Here, it is simplest to integrate by parts so that
\begin{equation}
\left. \sum_{n=0}^\infty I_2(n) t^n = -\frac{1}{\pi} \im\ G(x,t) \log \Gamma(\frac14 + \frac{i}{2} x) \right|_0^\infty - \frac{1}{\pi} \re \int_0^\infty dx\ G(x,t) \frac{-\frac12 \Gamma'(\frac14+\frac{i}{2} x)}{\Gamma(\frac14 + \frac{i}{2} x)} ~.
\end{equation}
We know that the boundary term $G_n(x) N(x)$ vanishes at the endpoints $x=0$ and $x=\infty$.
This does not imply that the boundary term vanishes for each of the summands in $N(x)$ separately as a detailed cancelation is in principle possible.
It turns out that it does vanish exactly for $I_2$.
At $x=0$, $G(0,t) \log\Gamma(\frac14)$ has zero imaginary part.
At $x=\infty$, the boundary term evaluates to
\begin{equation}
-\frac{t(1+t)}{2\pi(1-t)^3} \left( \frac{\log x}{x} - \frac{1+\log 2}{x} + {\cal O}(\frac{1}{x^2}) \right) \longrightarrow 0 ~.
\end{equation}
As we might expect from~\eqref{bdypiece}, the generating function for the square integers captures the $t$ dependence.

Using~\eqref{logDGamma}, we have that
\begin{align}
\nn
\sum_{n=0}^\infty I_2(n) t^n &= 
-\frac{1}{\pi} \re
\int_0^\infty dx\ \frac{4 t (1+t)}{4 (1-t)^3 x^2+(1+t)^2 (1-t)} \left[
\frac12 \gamma + \sum_{m=1}^\infty \left(
\frac{1}{- \frac32 + ix + 2m} - \frac{1}{2m} 
\right)
\right] \\
\nn
&=
\left[
-\frac{\gamma t}{2(1-t)^2} +  
\frac{t(1-2t)}{2(1-t)^2} \sum_{m=1}^\infty \frac{1}{m(1-2 t- 2 m (1-t))}
\right]
\\
&=
-\frac{\gamma t }{2 (1-t)^2} + \frac{t \left(\gamma + \psi\left(\frac{1}{2-2 t}\right) \right)}{2(1-t)^2} 
= 
\frac{t \psi\left(\frac{1}{2-2 t}\right)}{2(1-t)^2} 
=
t \frac{d}{dt} \log \Gamma\left(\frac{1}{2-2 t}\right) ~,
\end{align}
with $\psi(z) := (\log \Gamma(z))'$ the digamma function again.

To find an exact formula for $I_2(n)$, we expand about $t=0$:
\begin{eqnarray}
&& t \frac{d}{dt} \log \Gamma(z) = t \frac{dz}{dt} \frac{d\log\Gamma(z)}{dz} ~, \qquad z = \frac{1}{2-2t} \quad \Longrightarrow \quad t \frac{dz}{dt} = \frac12 \sum_{n=1}^\infty n t^n ~, \nn \\
&& \frac{d\log \Gamma(z)}{dz} = \sum_{n=0}^\infty \frac{1}{(2n)!} \psi^{(2n)}(\frac12) (z-\frac12)^{2n} + \sum_{n=1}^\infty (2^{2n}-1) \zeta(2n) (z-\frac12)^{2n-1} ~,
\end{eqnarray}
where $\psi^{(k)}(z) = \frac{d^k}{dz^k} \psi(z)$.
In fact, as $\psi^{(k)}(\frac12) = (-1)^{k+1} k! (2^{k+1}-1)\zeta(k+1)$, we deduce that
\begin{equation}
\frac{d\log \Gamma(z)}{dz} = \psi(\frac12) + \sum_{m=2}^\infty (-1)^m (2^m-1) \zeta(m) (z-\frac12)^{m-1} ~.
\end{equation}
The series expansion for $(z-\frac12)^{m-1}$ about $t=0$ tells us that
\begin{equation}
(z-\frac12)^{m-1} = \frac{t^{m-1}}{2^{m-1}} \sum_{\ell=0}^\infty \frac{t^\ell}{\ell!} \left( \prod_{k=0}^{\ell-1} (k+m-1) \right) ~.
\end{equation}
We also recall from before that $\psi(\frac12) = -\gamma - 2\log 2$.
Putting the pieces together,
\begin{equation}
t \frac{d}{dt} \log \Gamma(z) = -(\frac\gamma2 + \log 2) \sum_{n=1}^\infty n t^n + \sum_{n=1}^\infty \sum_{m=2}^\infty \sum_{\ell=0}^\infty n (-1)^m (1-2^{-m}) \zeta(m) \frac{1}{\ell!} \left( \prod_{k=0}^{\ell-1} (k+m-1) \right) t^{\ell+m+n-1} ~.
\end{equation}

We first of all note that the exponent of $t$ in the triple summation is at least $2$.
To extract the coefficient of $t^N$, where $N=\ell+m+n-1$, let us work term by term in $m$, which, since $n\ge 1$ and $\ell \ge 0$, has a maximum value of $N$.
Given $N$ and $m$, $n$ ranges from $1$ to $N-m+1$, and $\ell = N-m-n+1$ is fixed.
Using Pochhammer identities, we simplify the triple summation as follows:
\begin{eqnarray}
&& \sum_{N=2}^\infty \sum_{m=2}^N (-1)^m (1-2^{-m}) \zeta(m) \sum_{n=1}^{N-m+1} \frac{n}{(N-m-n+1)!} \left( \prod_{k=0}^{N-m-n} (k+m-1) \right) t^N \nn \\
&& \qquad = \sum_{N=2}^\infty \sum_{m=2}^N (-1)^m (1-2^{-m}) \zeta(m) \frac{N!}{m!(N-m)!} t^N ~.
\end{eqnarray}
We finally arrive at the explicit expression
\begin{equation}\label{alphan}
I_2(n) = -n(\frac\gamma2 + \log2) + \sum_{m=2}^n (-1)^m (1 - 2^{-m}) {n\choose m} \zeta(m) ~.
\end{equation}

In summary then, we have a relation between the $n$-th Li's coefficient with the integral of the oscillatory part.
This implies the following result.
\begin{corollary}
The Riemann hypothesis is equivalent to the Li coefficients being equal to the integral
\begin{eqnarray}
k_n &=& I_1(n) + I_2(n) - \frac{1}{\pi} \int_0^\infty dx\ G_n'(x) \arg \zeta(\frac12 + i x) ~, \nn
\end{eqnarray}
where the Li coefficient $k_n$ is defined in Proposition~\ref{li}, the constants $I_1(n)$ and $I_2(n)$ are defined, respectively, in~\eqref{I1} and~\eqref{alphan}, and the integration kernel $G'_n(x)$ is defined in terms of Chebyshev polynomials using~\eqref{Gn}.
\end{corollary}
We can readily check that for $n=1$, we recover the Volchkov criterion.

An immediate consequence of Li's criterion is then
\begin{corollary}
The Riemann hypothesis is equivalent to the positivity of the expression
\[
k_n = I_1(n) + I_2(n) + \frac{n}{\pi} 
\int_0^\infty dx\ 
\cU_{n-1}\left( \frac{x^2 - \frac14}{x^2 + \frac14} \right)
\frac{2 x \arg \zeta(\frac12 + i x) }{(\frac14 + x^2)^2}  > 0 ~,
\]
for all $n=1,2,3,\ldots$.
\end{corollary}

\subsection{An Integral Transform}\label{s:trans}
Having seen the inner workings on the types arguments leading to the Volchkov criterion, and of the relation to the integral representation of Li's coefficients, one is naturally led to whether integral kernels other than $(\frac14 + x^2)^{-1}$ would lead to nice statements.
Indeed, from a physical point of view, exponentials emerge more naturally, especially within the context of scattering amplitudes.
Now, one quickly sees that the most difficult integral is encountered when one integrates against the digamma function which is done by integrating against a rational function and performing an infinite sum.
There are indeed very few possibilities to allow both steps to be done exactly.
The selection of the kernel $G(x) = (\frac14 + T^2)^{-1}$ is a quite astute and special choice.

Nevertheless, this conducive choice can itself be written in terms of exponentials, either by a transform or a series of Laplace/Fourier type.
To do so, one invokes the Fourier transform
\begin{equation}
\frac{-2T}{(T^2+ \frac14)^2} 
= i \int_{-\infty}^\infty dx\ e^{i T x}\, x\, e^{-\frac12 |x|} \ ,
\label{ft}
\end{equation}
which, upon substitution into Proposition~\ref{vv}, supplies the following equivalent statement of the Riemann hypothesis:
\begin{equation}
e^{-2\pi i \gamma} =
\exp\left\{
- \frac12
\int_{-\infty}^\infty dx\ x\, e^{-\frac12|x|}\,  
\cF_x\left[
\arg \hat{A}(T)
\right]
\right\} \ , \label{inttr}
\end{equation}
where $\cF_x\left[
\arg \hat{A}(T)
\right]
=
\int_{-\infty}^\infty dT\  
e^{i T x}\, 
\arg \hat{A}(T)$ is the Fourier transform of the argument of the symmetrized Veneziano amplitude.
In recasting the Volchkov--Veneziano criterion in this way, we have made use of the fact that both~\eqref{ft} and $\arg \hat{A}(T)$ are odd functions of $T$.
This means that we can rewrite~\eqref{vv2} as an integral over the real axis.

\section{A Physical Picture}\label{physical}
\setall

As emphasized in the first section of this note, the Hilbert--P\'olya approach to the Riemann hypothesis posits the existence of a physical system whose spectrum is captured by the non-trivial Riemann zeros.
We have found an interesting relation between a particular physical system, to wit, string theory and a particular integral formulation of the Riemann hypothesis.
The spectrum of string theory can be read of from the Veneziano amplitude which knows only about the trivial Riemann zeros.
However, the phase of the Veneziano scattering amplitude can be used to relate string theory to the Riemann hypothesis through what we call the \textit{Volchkov--Veneziano criterion}.\footnote{According to~\cite{gannon}: ``In the 1962 International Congress of Mathematicians, I.~M.~Gel'fand remarked somewhat cryptically that there is an intriguing analogy between the scattering matrix of
quantum mechanics and zeta functions.'' The use of this intuition in the relation of scattering theory and automorphic forms has
led to some remarkable results~\cite{lax}. In some sense, our main intuition in this paper is that string theory provides the correct
quantum system for this relation between scattering theory and the Riemann zeta function. This claim could be in principle generalized to other L-functions as well.}

We note that the Volchkov--Veneziano criterion is reminiscent of the sum rules for the phase shift found in the physics literature~\cite{newton}. In particular, the well known Friedel sum rule~\cite{kittel} from solid state physics comes to mind when
thinking about the physical meaning of the Volchkov--Veneziano criterion. In this concluding section we wish to comment on the
physics and contemplate the possible further use of the picture we develop.
One might think that, from a physical point of view, our discussion is somewhat formal given the need to consider arbitrary complex momenta and the integration thereof. (However, note that complex momenta (and complex phase space) do appear in the context of
$PT$ quantum theory~\cite{bender}.)
Even including this caveat, the resulting picture is remarkably compelling as it suggests a new playground that bridges string theory and analytical number theory.

First, let us recall that in order to compute the four-point Veneziano amplitude from string theory, we need to insert four (tachyon) vertex operators on the Polyakov path integral on the disc (for an open string)~\cite{joe}. More precisely,
the four tachyon amplitude is given as (up to the factors of coupling and the delta function imposing the
momentum conservation~\cite{joe} as well as  the symmetrization of the momenta)
\begin{equation}
S_4 = \int_{-\infty}^{+\infty} d y_4\ \langle \prod_{j=1}^{3} : c^1(y_j) e^{i k_j X(y_j)}:: e^{i k_4 X(y_4)} : \rangle ~.
\end{equation}
Here, the expectation value of the inserted (and normal ordered) vertex operators $ {\cal O} \equiv e^{i k_j X(y_j)}$
($k_i$, or $k_1$, $k_2$, $k_3$, and $k_4$, are the momenta introduced in Section~\ref{secone} of our paper)
 and the $c$ ghost field, are computed (on the disc)
with the Polyakov worldsheet action for a string in the flat background 
$S_P (X)= \frac{1}{4 \pi \alpha'} \int d^2 \sigma\ \partial_a X^{\mu} \partial^a X_{\mu} $.
In general, a treatment of amplitudes in string theory and the $i\varepsilon$ prescription is subtle~\cite{wit}.

We may write
\begin{equation}
\langle {\cal O} (X) \rangle = \int D X\ {\cal O} (X) e^{- S_P (X)} = \int D X\ {\cal O} (X) e^{-  \frac{1}{4 \pi \alpha'} \int d^2 \sigma\ \partial_a X^{\mu} \partial^a X_{\mu}} ~.
\end{equation}
Next, we note that in order to obtain the Volchkov--Veneziano result, from a path integral point of view, we have to first integrate over the (free, \textit{i.e.}, non-interacting) worldsheet fields, thus obtaining the Veneziano amplitude~\cite{joe}; then we have to 
evaluate its argument; and finally, we need to integrate the argument of the Veneziano amplitude over the momenta, according to the measure featuring in the Volchkov--Veneziano criterion. More precisely we have
\begin{equation}
\int_{0}^\infty dT\ 
\frac{2 T}{(\frac14 + T^2)^2} \arg \left( \frac{\zeta(iT)}{\zeta(1-iT)}
{\int_{-\infty}^{+\infty} d y_4\ \langle \prod_{j=1}^{3} : c^1(y_j) e^{i k_j X(y_j)}:: e^{i k_4 X(y_4)} : \rangle  (T) } \right)
= 2\pi ( \gamma -3 ) ~.
\end{equation}
 
However, we can now also envision the opposite procedure.
In other words, we can first integrate over the momenta (and thus perform the integration over $T$ first) in the above explicit expression for the Volchkov--Veneziano criterion. This procedure would leave us with a pure path integral of an effective, yet interactive (and non-local) two dimensional theory of worldsheet fields on the disc.
We might call this effective two dimensional theory, the ``Riemann string.'' In other words instead of the Polyakov action 
we would have an effective non-local action over $X$
\begin{equation}
S_\mathrm{RS} = \frac{1}{4 \pi \alpha'} \int d^2 \sigma\ [\partial_a X^{\mu} \partial^a X_{\mu} +  F(X, \partial X, \partial^2 X,\ldots)] ~,
\end{equation}
where the non-local factor $ F(X, \partial X, \partial^2 X,\ldots)$ comes from the integration over the four momenta.

From this point of view the Volchkov--Veneziano criterion would become a criterion for the amplitude of this effective path integral of the Riemann string. However, the most delicate part of this formal procedure would be taking into account the argument of the
path integral (or more precisely, the argument of the generating functional of the Veneziano amplitude)!  In order to do this properly we would need some effective technical tool which computes the generating
functional for the two dimensional conformal field theory. One such technique could include the holographic
realization of Veneziano-type amplitudes~\cite{holo}
\begin{equation}
\langle e^{-\int d^2 \sigma k X}  \rangle \equiv \int D X e^{-\int d^2 \sigma k X} e^{-  \frac{1}{4 \pi \alpha'} \int d^2 \sigma \partial_a X^{\mu} \sigma \partial^a X_{\mu}}  = e^{ - S_H (p)} ~,
\end{equation}
where the $S_H(p)$ is the relevant holographic action~\cite{holo}, in terms of bulk variables $p$ whose ``boundary'' values would
correspond to the momenta $k$. After continuing the momenta to the Lorentzian signature, we could use such a holographic expression to evaluate the argument of the Veneziano amplitude.

What could be the physical meaning of this Riemann string? Unlike the canonical string theory, such a Riemann string theory might know about the non-trivial Riemann zeros. More precisely, instead of the canonical four-point amplitude (the Veneziano amplitude)
given by the Euler Beta function, $\frac{\Gamma(x) \Gamma (x)}{\Gamma(x+y)}$,
the Riemann string could have instead the appropriate zeta function analog, $\frac{\xi(x) \xi (x)}{\xi(x+y)}$.
This is, of course, purely conjectural at this point. However, 
further study of such an effective Riemann string path integral by various methods of string theory, including the holographic realization of Veneziano-type of amplitudes~\cite{holo}, the relation of Veneziano-like amplitudes and algebraic geometry~\cite{kholo}, the matrix model technology~\cite{matrixm} as well as the new view of
string theory in phase space~\cite{flm}, might open a new field of investigation relating the mathematics and physics of string theory to properties of prime numbers.

Particularly intriguing is the last string theory restatement of the Riemann hypothesis in terms of an integral 
transform of the argument of the Veneziano amplitude,~\eqref{inttr}.
If one could provide a holographic rewriting of the argument (or, alternatively, the logarithm) of the Veneziano amplitude in terms of some ``bulk'' (\textit{i.e.}, holographically dual) effective action, then the Riemann hypothesis would amount to a particular value (given by the proper regularization of the bulk term)
of the first moment of the Laplace transform of the bulk effective action, evaluated at fixed momentum (corresponding to the location of the critical line). Alternatively, the formula~\eqref{inttr} might have a nice rewriting in terms of the p-adic Veneziano amplitude, thus taking us back to the starting point of this paper.

The Riemann string might be very complicated to analyze. However, given the Freund--Witten representation of the Veneziano amplitude we could also imagine investigating the traditional string theory for the momenta which take values corresponding to the non-trivial Riemann zeros.
In this case, we would have to evaluate the usual Veneziano amplitude, \textit{i.e.}, the Euler Beta
(and thus first the Euler Gamma) function at the non-trivial Riemann zeros.
The natural question arises: Do we get Regge trajectories in this case as well?
If so, these Regge trajectories would not correspond to the usual string spectrum, but to some {\it dual} formulation of string theory (\textit{i.e.}, the Riemann string) which knows about the non-trivial Riemann zeros. %

In this paper, we make no claim to producing a solution to the Riemann hypothesis.
We believe that rephrasing it as a problem in physics renders it more tractable and constitutes an advance in its own right.
While string theory presents a model for quantum gravity, much of its considerable success to date has been in connecting physics to mathematics and linking different areas of mathematics to each other.
Indeed, many areas of string theory fall under the rubric of \textit{physical mathematics}~\cite{moore}.
The Riemann hypothesis resides at the nexus of number theory, analysis, and geometry.
It is therefore completely natural that string theory and the Riemann hypothesis couple to each other as well.

\section*{Acknowledgements}
We thank Lara Anderson, Thom Curtright, Peter Freund, James Gray, Seung-Joo Lee, Cyril Matti, Joan Sim\'on, and Mark Spradlin for discussions and communications. 
YHH is indebted to the Science and Technology Facilities Council, UK, for grant ST/J00037X/1, the Chinese Ministry of Education, for a Chang-Jiang Chair Professorship at NanKai University, and the city of Tian-Jin for a Qian-Ren Award.
YHH is also perpetually indebted to Merton College, Oxford for continuing to provide a quiet corner of Paradise for musing and contemplations.
VJ is supported by the South African Research Chairs Initiative of the Department of Science and Technology and the National Research Foundation and expresses gratitude to Queen Mary, University of London, McGill University, and the Perimeter Institute for hospitality.
DM is supported in part by the US Department of Energy under contract DE-FG02-13ER41917 and thanks the City University, London for hospitality.

\appendix
\setall
\section{Veneziano and Balazard--Saias--Yor}\label{ap:BSY}\setall
Taking the log absolute value of~\eqref{lim}, we have that
\begin{equation}
\log|A(\alpha^{-1}(-\frac12 + iT),\alpha^{-1}(-\frac12 + iT))| =
\log\left| 
\frac{\Gamma(i T)}{\Gamma(\frac12 - iT )} 
\pi^{\frac12-2iT} \right|
+ 2 \log | \zeta(\frac12 + i T) | - 2 \log | \zeta(\frac12 - i T) |
\end{equation}
Unfortunately, at this point because the zeta function is meromorphic,
\begin{equation}
| \zeta(\frac12 + i T) | = | \zeta(\frac12 - i T) | ~,
\end{equation}
so the interesting part of the above expression cancels out.
Nevertheless, let us proceed to arrive at a nice result.
Integrating both sides, after dividing by $\frac14 + T^2$, we have that
\begin{equation}
\int_{-\infty}^\infty 
\frac{dT}{\frac14 + T^2}
{\log\left|A(\alpha^{-1}(-\frac12 + iT),\alpha^{-1}(-\frac12 + iT))\right|}
= \int_{-\infty}^{\infty} \frac{dT}{\frac14+T^2}
\log\left|\frac{\Gamma(iT) \pi^{\frac12 - 2iT}}{\Gamma(\frac12 - iT)}\right|
\ . \label{absval}
\end{equation}
By recalling the duplication formula for the Gamma function~\eqref{dup} and the reflection formula~\eqref{reflection}, the absolute value in the integrand on the right hand side of~\eqref{absval} can be simplified to read
\begin{equation}
\log\left|\frac{\Gamma(iT) \pi^{\frac12 - 2iT}}{\Gamma(\frac12 - iT)}\right|
= \log (2 \cosh (\pi T) \left| \Gamma (2 i T)\right| ) \ .
\end{equation}
We then invoke the absolute value identity (see 8.332.1 in~\cite{gr})
\begin{equation}\label{absg}
\left| \Gamma (i y)\right| =\sqrt{\pi } \sqrt{\frac{\text{csch}(\pi 
   y)}{y}} ~, \qquad y\in \mathbb{R}
\end{equation}
to simplify further:
\begin{equation}
\mathrm{RHS} = \int_{0}^\infty  dT\  
\frac{\log \left(\frac{\pi  \coth (\pi  T)}{T}\right)}{T^2+\frac14}
\ .
\end{equation}
Now, employing the standard infinite product expressions for the hyperbolic trigonometric functions
\begin{equation}
\sinh(z) = z \prod_{n=1}^\infty \left(1 + \frac{z^2}{n^2\pi^2} \right) \ ,
\quad
\cosh(z) = \prod_{n=1}^\infty \left(1 + \frac{4z^2}{(2n-1)^2\pi^2} \right) \ ,
\end{equation}
we deduce that
\begin{equation}
\log \left(\frac{\pi  \coth (\pi  T)}{T}\right)
= -2 \log T + \sum_{n=1}^\infty \left[
\log \left(\frac{4 T^2}{(2 n-1)^2}+1\right) - 
\log \left(\frac{T^2}{n^2}+1\right) \right] \ .
\end{equation}
Whence, the right hand side of~\eqref{absval} becomes
\begin{align}
\nn
\mathrm{RHS} &=  \int_{0}^\infty  dT 
\left(
\frac{-2 \log T}{T^2 + \frac14} +
\sum_{n=1}^\infty \left[
\frac{\log \left(\frac{4 T^2}{(2 n-1)^2}+1\right)}{T^2 + \frac14} -
\frac{\log \left(\frac{T^2}{n^2}+1\right)}{T^2 + \frac14} \right]
\right) \\
\nn
&= \pi \log{4} + 
\pi \sum_{n=1}^\infty \log \left(\frac{16 n^4}{\left(1-4 n^2\right)^2}\right) \\
&= 2 \pi \log\pi \simeq 7.19255 ~,
\end{align}
which agrees perfectly with numerical integration.
To summarize
\begin{equation}
\int_{-\infty}^{\infty} \frac{dT}{\frac14+T^2}
\log\left|\frac{\Gamma(iT) \pi^{\frac12 - 2iT}}{\Gamma(\frac12 - iT)}\right|
= 2 \pi \log\pi \ .
\end{equation}
This is a nice exact integral.
The comparison to the Veneziano amplitude holds in the limit~\eqref{lim} independent of the Riemann hypothesis.

\section{A Subtle Phase}\label{ap:VV}
In this Appendix, we study the subtleties involved in case \textbf{A} described in \eqref{venlimit}.
Here, we use the ratio~\eqref{ratio} to eliminate one pair of zeta functions to arrive at
\begin{equation}\label{lim}
\tilde{A}(T) :=
A_4(\alpha^{-1}(-\frac12 + iT),\alpha^{-1}(-\frac12 + iT))
= \frac{\zeta(\frac12+iT)^2}{\zeta(\frac12-iT)^2}\frac{\Gamma(iT)}{\Gamma(\frac12-iT)}\pi^{\frac12-2iT} \ .
\end{equation}
We note that $\tilde{A}(T)$ and $\tilde{A}(-T)$ have the same real part, and the imaginary part differs by a sign.
We may equivalently express~\eqref{lim} as
\be
\tilde{A}(T) = \sqrt\pi\ \frac{\Gamma(\frac14-\frac{i}{2} T)^2}{\Gamma(\frac14+\frac{i}{2} T)^2}\frac{\Gamma(iT)}{\Gamma(\frac12-iT)} ~. \label{beforeprevious}
\ee
For later reference, we make note of the intermediate step in this identification, which also arises from application of~\eqref{ratio}:
\be
\frac{\Gamma(\frac14-\frac{i}{2}T)}{\Gamma(\frac14+\frac{i}{2}T)} = \frac{\zeta(\frac12+iT)}{\zeta(\frac12-iT)} \pi^{-iT} ~. \label{previous}
\ee

Let us consider the limit~\eqref{lim}, and test the complex argument.
That is, using $\arg \overline{z} = -\arg z$ and the additivity of the argument, we have that
\begin{equation}\label{limArg}
\arg \tilde{A}(T) = 4 \arg \zeta(\frac12+iT) + \arg \Gamma(i T) - \arg \Gamma(\frac12 - i T) - 2T \log\pi + 2\pi \tilde{n}(T) \ ,
\end{equation}
where $\tilde{n}(T)$ is an integer valued function.
Note that we are taking the argument to be principal and thus modulo $[-\pi, \pi)$, within the integral we shall shortly take; the constant adjustment of the argument to be within this interval is crucial for convergence.
The additive integer $\tilde{n}(T)$ is a guarantee of this principality.
We shall soon observe that $\tilde{n}(T)$ jumps discontinuously at the critical Riemann zeros.

Next, we invoke the nice exact expression that
\begin{equation}\label{gammaId}
\arg \Gamma(x + i y) = y \psi(x) + 
\sum_{n=0}^\infty \left( \frac{y}{x+n} - \arctan \frac{y}{x+n} \right)
\ ,
\end{equation}
where $x \in \IR_+, y \in \IR$ and avoiding the poles (\textit{i.e.}, $x+ i y \ne 0,-1,-2,\ldots$) and $\psi(x) = \Gamma'(x) / \Gamma(x)$ is the digamma function~\cite{wolfram}.
Whence, we have that
\begin{equation}\label{argdiff}
\arg\Gamma(\epsilon + i T) - \arg\Gamma(\frac12 - i T)
= T \psi(\epsilon) + T \psi(\frac12) +
\sum_{n=0}^\infty \left( 
\frac{T}{\epsilon+n} +  \frac{T}{\frac12+n} 
- \arctan \frac{T}{\epsilon+n}  - \arctan \frac{T}{\frac12+n} 
\right)
\ ,
\end{equation}
where we shall take $\epsilon \to 0$ at the end.
The reason we do this is that the digamma function has a pole at zero
\begin{equation}
\psi(\epsilon) = 
-\frac{1}{\epsilon }-\gamma +\frac{\pi ^2}{6} \epsilon +
\cO\left(\epsilon^2\right) ~,
\end{equation}
which will be seen to cancel precisely against a divergence in an infinite sum.
This is the same reason why $x$ is required to be positive in the identity~\eqref{gammaId}.

Applying~\eqref{argdiff} to~\eqref{limArg} and integrating against the kernel $\frac{2T}{(\frac14 + T^2)^2}$, we have that 
\begin{align}
\nn
&\int_0^\infty dT\ \frac{2 T \arg \tilde{A}(T)}{(\frac14 + T^2)^2} 
= 
4 \int_0^\infty dT\ \frac{2 T \arg \zeta(\frac12 + i T)}{(\frac14 + T^2)^2} + 
\\
\nn
&
+ \int_0^\infty dT\ \frac{2T\, 2\pi \tilde{n}(T)}{(\frac14 + T^2)^2} +
\int_0^\infty dT\ \frac{(2(\psi(\epsilon) + \psi(\frac12)) - 4\log\pi) T^2}{(\frac14 + T^2)^2} + 
\\
& + \sum_{n=0}^\infty
\int_0^\infty dT\ \frac{2 T}{(\frac14 + T^2)^2}
\left( 
\frac{T}{\epsilon+n} +  \frac{T}{\frac12+n} 
  - \arctan \frac{T}{\epsilon+n}  - \arctan \frac{T}{\frac12+n} 
\right). \label{theint}
\end{align}

Thus, we have that $\mathrm{LHS} = \mathrm{RHS}$, where
\begin{align}
\nn
&\mathrm{LHS} :=\int_0^\infty dT\ \frac{2 T \arg \tilde{A}(T)}{(\frac14 + T^2)^2}-4 \int_0^\infty dT\ \frac{2 T \arg \zeta(\frac12 + i T)}{(\frac14 + T^2)^2}-2\pi \int_0^\infty dT\ \frac{2T\, \tilde{n}(T)}{(\frac14 + T^2)^2}
\ ,\\
\nn
\qquad
& \mathrm{RHS} := 
\int_0^\infty dT\ 
\frac{(2(\psi(\epsilon) + \psi(\frac12)) - 4\log\pi) T^2}{(\frac14 + T^2)^2} + 
\\
\qquad
& + \sum_{n=0}^\infty
\int_0^\infty dT\ \frac{2 T}{(\frac14 + T^2)^2}
\left( \frac{T}{\epsilon+n} +  \frac{T}{\frac12+n} 
  - \arctan \frac{T}{\epsilon+n}  - \arctan \frac{T}{\frac12+n} \right)
\ . \label{argint}
\end{align}
The integrals on the right hand side are elementary, \textit{viz.}, (for $x\in\IR_+, n\in\IZ_+$)
\begin{equation}\label{elemint}
\int_0^\infty dT\ \frac{2 T^2}{(\frac14 + T^2)^2} = \pi  \ ,
\qquad
\int_0^\infty dT\ 
\frac{2 T \arctan
 \left(\frac{T}{x+n}\right)}{\left(T^2+\frac{1}{4}\right)^2}
= \frac{\pi}{\frac12 + x + n}
\ ;
\end{equation} 
whence, we discover
\begin{equation}
\mathrm{RHS} = 
- 2\pi \log \pi 
+ \pi \psi(\frac12) 
+ \pi \left[
\psi(\epsilon) + \sum_{n=0}^\infty \left(
\frac{1}{\epsilon+n} + \frac{1}{\frac12+n} 
- \frac{1}{\frac12+\epsilon+n} - \frac{1}{1+n} \right)
\right] \ .
\end{equation}
Here, crucially, we have isolated the $\psi(\epsilon)$ with the divergent sum, which, in combination, cancels exactly since
\begin{equation}
\sum_{n=0}^\infty \left(
\frac{1}{\epsilon+n} + \frac{1}{\frac12+n} 
- \frac{1}{\frac12+\epsilon+n} - \frac{1}{1+n} \right)
= 
-\gamma - \psi(\frac12) + \psi(\frac12 + \epsilon) - \psi(\epsilon) \ .
\end{equation}
We will present numerical evidence of support for our manipulations in Appendix~\ref{ap:VV}.
Finally, $\psi(\frac12) = -\gamma - 2 \log 2$.

To evaluate the LHS in~\eqref{argint} we begin with~\eqref{previous}, which tells us that
\be
2\arg \Gamma(\frac14-\frac{i}{2} T) = 2\arg \zeta(\frac12+iT) - T\log\pi + 2\pi n(T) ~,
\ee
for some integer $n$ dependent upon $T$. Thus,
\be
n(T) = \frac{1}{2\pi} \left( 2\im\log \Gamma(\frac14-\frac{i}{2}T) - 2\im \log \zeta(\frac12+iT) + T\log\pi \right) ~.
\ee
To disentangle what $n(T)$ is let us recall that the number of zeros $\rho = \sigma + i T$ in the critical strip in the upper half plane with $T<x \in \IR$ is given by an exact formula (\textit{cf}.~Theorem~9.1 of Titchmarch or Section~6.6 of Edwards~\cite{edwards}):
\begin{equation}
N(x) = 
1 - \frac{\log\pi}{2\pi} x + \frac{1}{\pi} \im \log \Gamma(\frac14 + \frac{i}{2} x) + \frac{1}{\pi} \im \log \zeta(\frac12 + i x) \ . \label{countingfn}
\end{equation}
Here, $x$ is itself not the imaginary part of a root.
Traditionally, one defines the auxiliary function $\vartheta(x) := \im \log \Gamma(\frac14 + \frac{i}{2} x) - \frac{x}{2} \log \pi$ so that
\begin{equation}
N(x) = \frac{1}{\pi} \vartheta(x) + 1 + 
\frac{1}{\pi} \im \log \zeta(\frac12 + i x) ~,
\end{equation}
and the first term is an {\it average part} which is easy to control while the fluctuating piece, proportional to $\arg \zeta(\frac12 + i x)$, contains the non-trivial information about the Riemann zeros.

We recognize by inspection of the above expressions that $n(T) = 1-N(T)$.
Comparing~\eqref{lim},~\eqref{beforeprevious}, and~\eqref{previous}, we conclude as well that $\tilde{n}(T) = 2n(T)$.
The argument of the zeta function that appears in the LHS of~\eqref{argint} exactly cancels against the argument of the zeta function that appears in $\tilde{n}(T)$.
Thus, because of this subtle conspiracy, the Veneziano amplitude in case \textbf{A} does not contain information that encodes the Riemann hypothesis via the Volchkov criterion.

\subsection{A Numerical Check}
In evaluating~\eqref{theint}, we have determined that
\begin{eqnarray}
\int_0^\infty dT\ \frac{2T}{(\frac14+T^2)^2} \left( \arg \Gamma(\epsilon + iT) - \arg \Gamma(\frac12 - iT) \right) &=& \pi \psi(\frac12) - \pi \gamma = -2\pi ( \gamma + \log 2 ) \nonumber \\
&\simeq& -7.981925165390411\ldots ~.\label{exact}
\end{eqnarray}
Let us verify this result numerically.

We may simplify the integrand using the reflection formula~\eqref{reflection} and the duplication formula~\eqref{dup}:
\begin{equation}
\arg\Gamma(\epsilon + i T) - \arg\Gamma(\frac12 - i T) =
\arg\left(2^{-2 i T} \Gamma(2 i T) \right) 
= -T \log 4 + \arg \Gamma(2 i T)
\ .
\end{equation}
The first term integrates easily:
\begin{equation}
\int_0^\infty dT\ \frac{2 T}{(\frac14 + T^2)^2} 
\left(
-T\log 4
\right) = -\pi \log 4 \ , \label{partial}
\end{equation}
while the second summand gives us a non-trivial integral.

In order to check against numerical integration, let us use the following simple rewriting of the principal argument function 
\begin{equation}
\arg z = -i \left(\log z - \log |z| \right) ~, \qquad z \in \IC \ .
\end{equation}
The reason for this is purely computational.
In \texttt{Mathematica} the jumps in $\arg \Gamma(z)$ renders numerical treatment ineffective in precision, whereas the function $\log\Gamma(z)$ is far better to use (\textit{viz.}, use the {\sf \mbox{LogGamma[~]}} rather than the {\sf \mbox{Log[Gamma[~]]}} command).
Using the expression quoted in~\eqref{absg} to simplify $\log|\Gamma(z)|$, we have that
\begin{equation}
\arg \Gamma(2 i T) = -i(\log\Gamma(z) - \log |\Gamma(z)|)
= -i\log\Gamma(2 i T) 
+ \frac{i}{2} \log \frac{\pi \text{csch}(2 \pi T)}{2 T} \ .
\end{equation}

Happily, \texttt{Mathematica} can now perform this numerical integration without complaint and gives the answer
\begin{equation}
\int_0^\infty dT\ \frac{2 T}{(\frac14 + T^2)^2} 
\left(
-i\log\Gamma(2 i T) 
+ \frac{i}{2} \log \frac{\pi \text{csch}(2 \pi T)}{2 T} 
\right) \simeq -3.626752984783235\ldots ~.
\end{equation}
Adding this to~\eqref{partial} yields
\begin{equation}
-\pi \log 4 -3.6267529847832356\ldots \simeq -7.981925165390439\ldots \ .\label{numeric}
\end{equation}
As~\eqref{exact} and~\eqref{numeric} agree to $13$ decimal places, we have confidence in our manipulations.

\section{The Virasoro--Shapiro Amplitude}\label{ap:VS}

The Virasoro--Shapiro amplitude describes the scattering amplitude of four closed string tachyons~\cite{vs}.
In what follows, we set $\alpha' = \frac12$ and adopt the conventions of~\cite{gsw}.
Starting from
\be
A_4^\mathrm{closed}(s,t,u) = \frac{\Gamma(a)\Gamma(b)\Gamma(c)}{\Gamma(a+b)\Gamma(b+c)\Gamma(c+a)} ~,
\label{closed}
\ee
where
\bea
&& a=-\frac12 \alpha(s) ~, \qquad b=-\frac12 \alpha(t) ~, \qquad c=-\frac12 \alpha(u) ~, \\
&& \alpha(x) = 2+\frac14 x ~, \qquad s+t+u=\sum_{i=1}^4 m_i^2 = -32 ~,
\eea
after a trivial amount of algebra, we find
\be
A_4^\mathrm{closed}(s,t,u) = \prod_{x=s,t,u} \frac{\Gamma(-1-\frac18 x)}{\Gamma(2+\frac18 x)} = \prod_{x=s,t,u} \frac{\Gamma(z(x))}{\Gamma(1-z(x))} ~,
\ee
with $z(x):=-1-\frac18 x = -\frac12\alpha(x)$.

In the open string sector, for a slightly different Regge trajectory $\alpha(x)$, we previously expressed the symmetrized Veneziano amplitude as
\be
A_4^\mathrm{open}(s,t,u) = \sqrt\pi \prod_{x=s,t,u} \frac{\Gamma(z(x))}{\Gamma(\frac12-z(x))} = \prod_{x=s,t,u} \frac{\zeta(1-z(x))}{\zeta(z(x))} ~.
\label{open}
\ee
(See Proposition~\ref{FW} and~\eqref{vene1}.)
Here,
\be
z(x) = -\frac12 \alpha(x) ~, \qquad \alpha(x) = 1+\frac12 x ~, \qquad s+t+u = -8 ~.
\ee
Structurally, as Freund and Witten already observe~\cite{Freund:1987ck}, the Veneziano amplitude in the middle expression in~\eqref{open} is almost the same as the Virasoro--Shapiro amplitude for four closed string tachyons.

Let us look for a nice simplification of $A_4^\mathrm{closed}(s,t,u)$ in terms of zeta functions.
From the functional identity~\eqref{funcid}, we have
\bea
\Gamma(1-z) &=& 2^{-z} \pi^{1-z} \frac{\zeta(z)}{\zeta(1-z)} \frac{1}{\sin( \frac\pi2 z )} ~, \nn \\
\Gamma(z) &=& 2^{-1+z} \pi^z \frac{\zeta(1-z)}{\zeta(z)} \frac{1}{\sin( \frac\pi2(1-z) )} ~.
\eea
In the second line, we have sent $z\to 1-z$.
Taking the ratio and using $s+t+u=-32$, we find that
\be
A_4^\mathrm{closed}(s,t,u) = \frac{1}{2\pi} \prod_{x=s,t,u} \left( \frac{\zeta(1-z(x))}{\zeta(z(x))} \right)^2 \tan(\frac\pi2 z(x)) ~.
\ee

As the Regge trajectories are different for the closed bosonic string and the open bosonic string, to compare the Virasoro--Shapiro amplitude to the Veneziano amplitude, define $\hat{x} = \frac14 x$.
The hat will emphasize the open string variables that we have used in the main text.
We express the argument of the zeta functions $z(x) = -1-\frac18x = -\frac12 \alpha(x) = -\hat\alpha(\hat{x})$.
Using the expression for the Veneziano amplitude from Proposition~1, we have
\be
A_4^\mathrm{closed}(s,t,u) = -\frac{\left(A_4^\mathrm{open}(\hat{s},\hat{t},\hat{u})\right)^2}{2\pi} \prod_{x=s,t,u} \tan(\frac\pi2 \hat\alpha(\hat{x})) ~.
\label{clop}
\ee
This is the observation of Kawai--Lewellen--Tye~\cite{klt}.

Knowing this, after straightforward algebra, we find that
\be\label{opcl}
\arg A_4^\mathrm{open}(T) = -\frac\pi4 + \frac12 \tan^{-1}( \sinh(\pi T) ) + \frac12 \arg A_4^\mathrm{closed}(T) ~.
\ee
As in Proposition~\ref{vv}, we have evaluated in the limit~\eqref{newlimit} where
\be
\hat\alpha(\hat{s}) = -iT ~, \quad \hat\alpha(\hat{t}\,) = -\frac12 ~, \quad T\in \mathbb{R} ~, \quad \mathrm{with}\ \hat\alpha(\hat{s}) + \hat\alpha(\hat{t}\,) + \hat\alpha(\hat{u}) = -1 ~.
\ee
We may now substitute~\eref{opcl} into Proposition~\ref{vv} to express the Volchkov criterion for the Riemann hypothesis in terms of the closed string Virasoro--Shapiro amplitude.

\section{Representations of the Li Coefficients}\label{ap:Li}
For completeness, we sketch the proofs of the two equivalent representations of the Li coefficients in the right hand side of~\eqref{box}, following largely the original paper~\cite{li}, but being more careful with precise factors. These should supplant our integral representation from~\cite{us}.

Define $\varphi(z) = \xi(\frac{1}{1-z})$, which is also equal to $\xi(\frac{z}{z-1})$ by the functional equation~\eqref{func}, then the Hadamard product~\eqref{hadamard} gives us
\begin{equation}
\varphi(z) = \frac12 \prod_{\rho \in \mathfrak{S}} 
\left(1 - \frac{z/(z-1)}{\rho} \right) = \frac12 \prod_{\rho \in \mathfrak{S}} 
\frac{1 - \left(1 - \frac{1}{\rho} \right)z }{1-z} \ .
\end{equation}

Hence, we have that (for $|z| < \frac14$, say, which renders the following function analytic)
\begin{equation}
\frac{\varphi'(z)}{\varphi(z)} = \frac{d}{dz} \log \varphi(z) = 
\sum_{\rho \in \mathfrak{S}} 
\frac{d}{dz} \log \frac{1 - \left(1 - \frac{1}{\rho} \right)z }{1-z}
=
\sum_{n=1}^\infty k_{n+1} z^n \ ; \qquad
k_n := \sum_{\rho \in \mathfrak{S}} \left[
1 - \left(1 - \frac{1}{\rho}\right)^n
\right] \ ,
\end{equation}
where in the last step we used the elementary fact that $\log(\frac{1 - A z}{1 - z}) = \sum_{n=0}^\infty \frac{1}{n}(1 - A^n) z^n$.
This furnishes the first definition of the Li coefficients and obviously generalizing the constant $B$ in~\eqref{AB} which corresponds to $n=1$.

Next, we have that
\begin{align}\nn
\frac{1}{(n-1)!} \frac{d^n}{dz^n} \left[
z^{n-1} \log \xi(z)
\right]_{s=1} &= 
\frac{1}{(n-1)!} \sum_{k=0}^n 
{n \choose k}
\left[
\frac{d^{k}}{dz^{k}} z^{n-1} \frac{d^{n-k}}{dz^{n-k}} 
\sum_{\rho \in \mathfrak{S}} \log \left( 1 - \frac{z}{\rho} \right)
\right]_{z=1}
\\
\nn
&= 
\frac{1}{(n-1)!} \sum_{k=0}^n 
\sum_{\rho \in \mathfrak{S}} 
{n \choose k}
\left[
\frac{(n-1)!}{(n-k-1)!} z^{n-k-1}
\frac{- (n-k-1)!}{(\rho - z)^{n-k}}
\right]_{z=1}
\\
& =
- \sum_{\rho \in \mathfrak{S}} 
\sum_{k=0}^n {n \choose k} (\rho - 1)^{k-n} 
= \sum_{\rho \in \mathfrak{S}} \left[
1 - \left(1 - \frac{1}{\rho}\right)^n \right] \ ,
\end{align}
and we thus see that the left hand side is another convenient representation for $k_n$.

\end{document}